\documentclass[12pt,preprint,useAMS,usenatbibi,twocolumn]{emulateapj}
\usepackage{latexsym}
\usepackage{amssymb}
\usepackage{amsmath}
\usepackage{epsfig}
\usepackage{appendix}
\singlespace
\def\kms{\,{\rm km\,s^{-1}}}

\def\msun{\,{\rm M_\odot}}

\def\sfr{\,{\rm M_\odot\,yr^{-1}}}
\def\spose#1{\hbox to 0pt{#1\hss}}
\def\lta{\mathrel{\spose{\lower 3pt\hbox{$\mathchar"218$}}
     \raise 2.0pt\hbox{$\mathchar"13C$}}}
\def\gta{\mathrel{\spose{\lower 3pt\hbox{$\mathchar"218$}}
     \raise 2.0pt\hbox{$\mathchar"13E$}}}
\submitted{Draft Version}

\begin{document}
\title{Pseudobulge Formation as a Dynamical Rather than a Secular Process}

\author{Javiera Guedes\altaffilmark{1}, Lucio Mayer\altaffilmark{1,2}, Marcella Carollo\altaffilmark{1}, \& Piero Madau\altaffilmark{3}}
\altaffiltext{1}{Institute for Astronomy, ETH Z\"urich, Wolgang-Pauli-Strasse 27, 8093 Zurich, Switzerland.}
\altaffiltext{2}{Institute for Theoretical Physics, University of Z\"urich, Winterthurerstrasse 190, CH-9057 Zurich, Switzerland.}
\altaffiltext{3}{Department of Astronomy \& Astrophysics, University of California, 1156 High Street, Santa Cruz, CA 95064.}
 
\begin{abstract}

We investigate the formation and evolution of the pseudobulge in ``Eris", a high-resolution $N$-body + smoothed particle hydrodynamics (SPH) cosmological simulation that successfully 
reproduces a Milky Way-like massive late-type spiral in a cold dark matter ($\Lambda$CDM) universe. At the present epoch, Eris has a virial mass 
$M_{\rm vir}\simeq8\times10^{11}\msun$, a photometric stellar mass $M_{*}=3.2\times10^{10}\msun$, a bulge-to-total ratio $B/T = 0.26$, and a weak nuclear bar. We find that the bulk of the pseudobulge forms quickly at high redshift via a combination of non-axisymmetric
disk instabilities and tidal interactions or mergers both occurring on dynamical timescales,
not through slow secular processes at lower redshift. Its subsequent evolution is 
not strictly secular either, and is closely intertwined with the evolution of the stellar bar. In fact, the structure that we recognize as a pseudobulge today evolves from a stellar bar that formed at high redshift,  was destroyed by minor mergers at $z \sim 3$, reformed shortly after, and weakened again following a steady gas inflow at $z\lta 1$. The gradual dissolution of the bar ensues
at $z\sim1$ and continues until the present without increasing the stellar velocity dispersion in the inner regions.  In this scenario the pseudobulge is not a separate component from the inner disk in terms of formation path, rather it is the first 
step in the inside-out formation of the baryonic disk, in agreement with the fact that pseudobulges of massive spiral galaxies have typically a dominant old
stellar population. If our simulations do indeed 
reproduce 
the formation mechanisms of massive spirals, then the progenitors of late-type galaxies should have strong bars and small photometric pseudobulges at high redshift. 

\end{abstract}


%%%%%%%%%%%%%%%%%%%%%%%%%%%%%%%%%%
% INTRODUCTION
%%%%%%%%%%%%%%%%%%%%%%%%%%%%%%%%%%

\section{Introduction}

The formation of realistic galaxies in cosmological simulations has been a challenge over the past decade. The massive spiral galaxies formed in these simulations have traditionally been too centrally 
concentrated, with small disks and large bulges, a problem associated with the catastrophic loss of angular momentum \citep{navarro00,scannapieco12}. Recently, owing to the increase in 
numerical resolution and the improvement of sub-grid recipes for star formation and feedback, several groups have reported successful simulations in $\Lambda$CDM of disk systems 
over a range of masses, from field dwarfs \citep{governato10} to massive spirals \citep{piontek11,guedes11,brook12,Okamoto12}, to ellipticals \citep{johansson12}. The increasingly better resolution of these types of simulations is now allowing us to study the detailed mechanisms involved the formation of galaxies and their different components. In the following, we focus on the formation of the bright, central component of disk galaxies.

Possible scenarios for the formation of galactic bulges include early build-up via mergers at high redshift, secular evolution due to the presence of a bar 
slowly bringing gas to the center and turning into a spheroid via the buckling instability or resonant thickening \citep[e.g.][]{raha91,debattista04, debattista05}, and fragmentation 
via gravitational instability producing giant gas clumps that spiral-in towards the center via dynamical friction \citep[e.g.][]{noguchi1998, Immeli04, bournaud07}. These distinct mechanisms 
lead to the formation of galactic bulges of different characteristics.  ``Classical bulges" are thought to form via mergers \citep{naab06,hopkins10}, which can rapidly produce 
spheroids by violently destroying disks, and are characterized by a steep increase in density towards the galaxy center. These bulges are best fit by high S\'ersic index ($n>2$) 
profiles, resembling the surface brightness distribution of elliptical galaxies. ``Pseudobulges" are observed instead to have disk-like density profiles \citep{carollo97,carollo98a,carollo98b} 
and kinematics \citep{kormendy93}, low S\'ersic index ($n<2$) profiles, and are thought to originate from disk material via secular evolution \citep[see a review by][]{kormendy04} 
induced by spiral structure or bars \citep{combes81,pfenniger90,combes93,debattista04,athanassoula05}. While in general pseudobulges have younger stellar populations than classical bulges, massive late-type spirals tend to have old pseudobulges \citep{carollo2007}.

Traditionally, detailed studies of secular evolution have been confined to idealized isolated galaxy models, which by construction neglect the effect of cosmological accretion of 
matter, external perturbations by infalling satellites, and the fact that the disk itself can grow and be destroyed several times as a  result of mergers \citep[e.g.][]{robertson06,
governato09}. Notable exceptions are the work of \cite{curir06,curir07} who studied bar instabilities in cosmological simulations at relatively low mass resolution (a factor of 17 lower than Eris), and that of \cite{Okamoto12} who does not focus however on the interplay between the stellar bar and the formation of the pseudobulge, as we do here. Studies of isolated galaxy models have shown that the proper modeling of shocks that lead to the accumulation of gas and the subsequent formation of stars at Linblad resonances is necessary for the formation of realistic spiral structure, and that the amount of gas inflow is strongly dependent on the gravitational torques exerted by stellar bars. Therefore resolving the bar is crucial to studying the formation of the bulge. High gas concentrations can lead to the destruction of the bar \citep{friedli93a} due to the increase of chaotic stellar orbits \citep{hasan90, hasan93}, generate nuclear bars \citep{friedli93b}, fuel active galactic nuclei activity \citep{phinney94}, and dissolve bars into bulges \citep{norman96}. The amount of gas in the disk also influences the pattern speed of the bar by increasing the precessing frequency $\Omega-k/2$ \citep{friedli93a}, due to a steady transfer of angular momentum from the bar to disk stars \citep{sellwood81}. Bars may also trigger density waves in the dark matter distribution \citep{weinberg07} that can enhance and displace the dark matter annihilation signal from the centers of galaxies \citep{kuhlen12}.

In this paper, we use the  Eris simulation \citep{guedes11} to study the origin and evolution of the bulge component of a Milky Way-like late-type system, and its relation to the 
assembly of the disk, with unprecedented detail. Eris appears to be the first cosmological hydrodynamic simulation in which the galaxy structural properties, the mass budget in the 
various components, and the scaling relations between mass and luminosity are all consistent with a host of observational constraints. 
Because of its high resolution and remarkable agreement with the observations, Eris provides then an excellent laboratory for 
understanding the mechanisms that lead to the formation of a pseudobulge in a single massive spiral whose internal structure is well resolved.

%%%%%%%%%%%%%%%%%%%%%%%%%%%%%%%%%%
% THE ERIS SIMULATION
%%%%%%%%%%%%%%%%%%%%%%%%%%%%%%%%%%
\section{The Eris Simulation}

%%%%%%%%%%%%%%%%%%%%%%%%%%%%%%%%%%
% SIMULATION SETUP
%%%%%%%%%%%%%%%%%%%%%%%%%%%%%%%%%%
\subsection{Simulation setup}

Eris follows the formation of a $M_{\rm vir}=7.9\times 10^{11}\msun$ galaxy halo from $z=90$ to the present epoch in a {\it Wilkinson Microwave Anisotropy Probe} 3-year cosmology.
This ``zoom-in" simulation was performed using the $N$-body + smoothed particles hydrodynamics (SPH) code {\tt GASOLINE} \citep{wadsley04}. The version of the code used for Eris 
includes Compton cooling, atomic cooling,  metallicity-dependent radiative cooling at low temperatures \citep{mashchenko06}, a prescription for blastwave supernova 
feedback \citep{stinson06}, and the ionizing effect of a uniform UV background using a modified version of the \citet{haardt96} spectrum. The target halo was selected to 
have a quiet merger history  with no major mergers (defined as mass ratio $\ge 1/10$) at redshift $z<3$ (four minor mergers occur at $z\simeq 3$, and two more at $z\simeq 1$ and $z\simeq 0.6$). 
The high resolution region was resampled with 13 million dark matter particles and an equal number of gas particles, for a mass resolution of $m_{\rm DM}=9.8\times 10^4\,\msun$ 
and $m_{\rm SPH}=2\times 10^4\,\msun$. The gravitational softening length was fixed to 120  physical pc for all particle species from $z=9$ to the present time, 
and evolved as $1/(1+z)$ at earlier times. Star particles form in cold gas that reaches a density threshold of $n_{\rm SF}=5$ atoms cm$^{-3}$ at the rate 
$d\rho_*/dt = 0.1 \rho_{\rm gas}/t_{\rm dyn}$, where $\rho_*$ and $\rho_{\rm gas}$ are the stellar and gas densities, and $t_{\rm dyn}$ is the local dynamical time. 
They are created stochastically with an initial mass $m_*=6\times 10^3\,\msun$ distributed following a \citet{kroupa93} initial mass function. 
Supernova explosions deposit an energy of $8\times 10^{50}\,$ergs and metals into the nearest neighbor gas particles, and the heated gas has its cooling shut off following
\citet{stinson06}. The use of a high threshold for star formation has the effect of increasing the efficiency of supernovae feedback through the injection of energy in localized 
high-density regions. This results in the formation of an inhomogeneous interstellar medium \citep{governato10,guedes11} and the preferential removal of low angular momentum 
material \citep{brook10}.

\begin{figure*}[h!]
\vspace{-1.5cm}
\hspace{-1cm}
\includegraphics[width=0.95\textwidth]{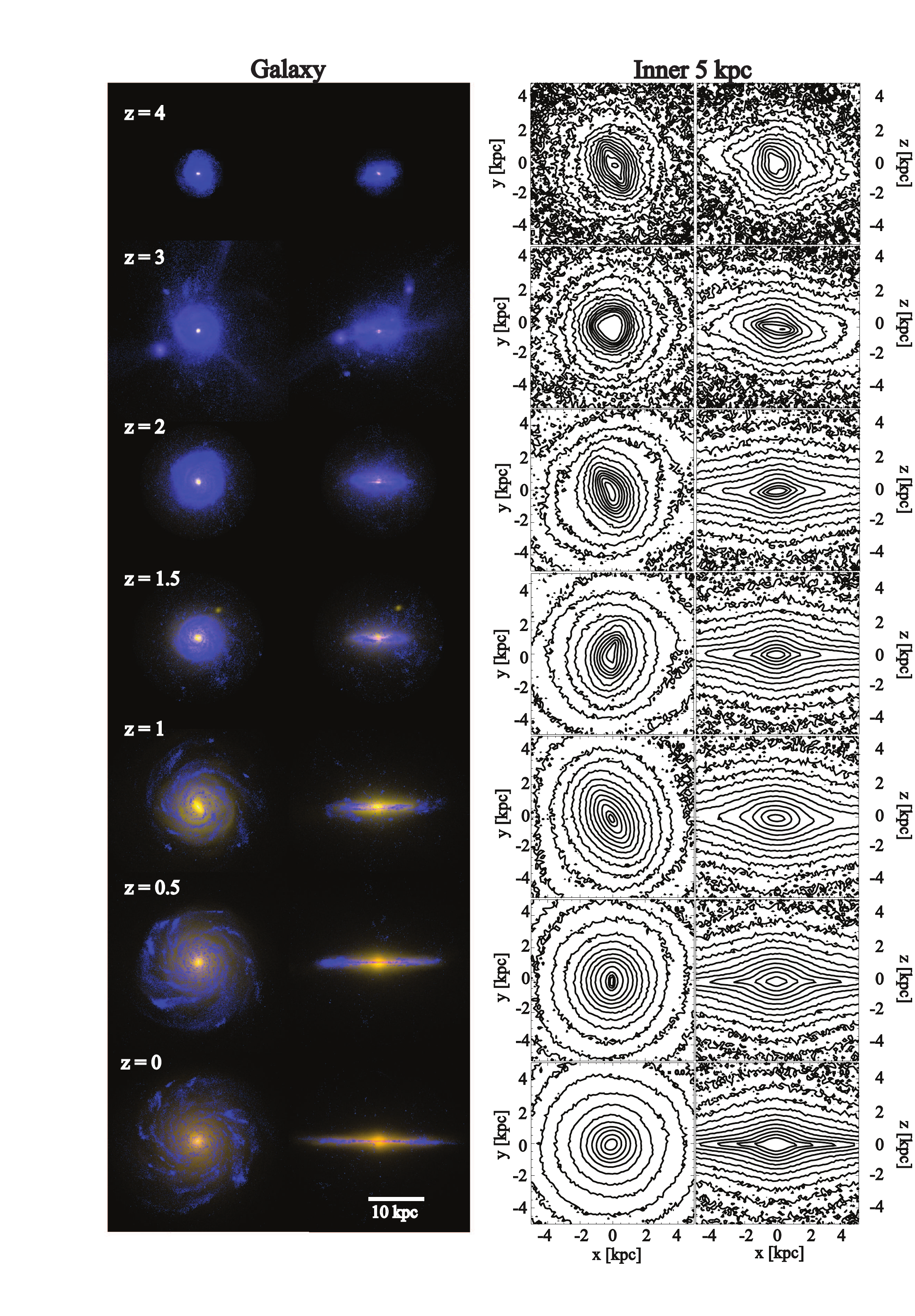}
\caption{Evolution of Eris from $z=4$ {\it (top)} to $z=0$ {\it (bottom)}. {\it Left panel}: composite $i$-band, $V$, and far UV rest-frame images generated using the {\tt SUNRISE} 
radiative transfer code \citep{jonsson06}. {\it Right panel}: Stellar surface mass density contours of the inner 5 comoving kpc of the galaxy, in face-on and edge-on projections through 10 
kpc slices. The bar forms together with the proto-disk at $z\sim5$, is destroyed by minor mergers at $z\sim3$, reforms soon afterwards, and finally weakens from $z\lta 1$ to the present.
}
\vspace{+0.cm}
\label{evolution}
\end{figure*}

% %%%%%%%%%%%%%%%%%%%%%%%%%%%%%%%%%%
% ERIS TODAY
%%%%%%%%%%%%%%%%%%%%%%%%%%%%%%%%%%%
\subsection{A realistic late-type massive spiral}

At $z=0$, Eris is a late-type spiral galaxy of virial mass $M_{\rm vir}=7.9\times10^{11} \msun$ and virial radius $R_{\rm vir}=239$ kpc, resolved with $N_{\rm DM}=7\times10^6$, 
$N_{\rm gas}=3\times10^6$, $N_{\rm *}=8.6\times10^6$ dark matter, gas, and star particles, respectively. To apply photometric analysis techniques to the simulation data, 
we use the radiative transfer code {\tt SUNRISE} \citep{jonsson06} in post-processing to generate mock images of the galaxy in a range of filters. Eris' $i$-band absolute magnitude 
and total photometric stellar mass are $M_i=-21.7$ and $M_{*}=3.2\times10^{10}\msun$, respectively. A two-dimensional decomposition of the SDSS $i$-band image using the {\tt GALFIT} 
package \citep{peng02} shows that Eris has an extended disk of exponential scale length $R_d=2.5$ kpc, a pseudobulge with S\'ersic index $n=1.4$, and a bulge-to-disk ratio B/D=0.35
\citep{guedes11}. The present-day $u-r = 1.52$ mag color of Eris is consistent with the colors of galaxies that host pseudobulges \citep{drory07}. Eris's surface brightness profile 
shows a downbending break at about five disk scale lengths, as observed in many nearby spiral galaxies \citep{pohlen06,bakos08}. The break radius $r_{\rm br}=10.7$ kpc, the 
extrapolated central surface brightness $\mu_{\rm in}=19.9$ $i$-mag arcsec$^{-2}$, and the break surface brightness $\mu_{\rm br}=23.5$ $i$-mag arcsec$^{-2}$ are all in 
excellent agreement with the observations \citep{pohlen06}. 

Eris falls on the Tully-Fisher relation as measured by \cite{pizagno07}, on the locus of the $\Sigma_{\rm SFR}$-$\Sigma_{\rm HI}$ plane occupied by nearby spiral galaxies \citep{bigiel08}, 
and on the stellar mass-halo mass relation at $z=0$ \citep{behroozi10}. Its rotation curve is consistent with observations of blue horizontal-branch halo stars in the
{\it Sloan Digital Sky Survey} \citep{xue08}. Its stellar disk is currently forming stars at a rate, $1.1\sfr$, comparable to the value inferred for the Milky Way \citep{robitaille10}. 
Its cold gas reservoir, $1.9\times10^9\msun$, is in good agreement with values inferred for the Milky Way \citep{Nakanishi03}. Cold gas
extends to six stellar disk scale lengths, and is pierced by kpc-size holes produced by SN explosions as observed in nearby spirals \citep{boomsma08}. 
%The hot ($T>3\times10^5$ K) halo gas follows a density profile of slope -1.13 to 100 kpc, shallower than the best-fit Navarro-Frenk-White dark matter profile \citep{navarro96}: 
%its dispersion measure (the integrated electron number density along the line of sight) out to 50 kpc (the distance of the Large Magellanic Cloud) is DM$=62\pm3$ cm$^{-3}$ pc, 
%consistent with Galactic pulsar measurements \citep{anderson10}. 

 Twin simulations of the Eris galaxy that include metal-dependent radiative cooling at high temperature ($T > 10^4$ K) and explicit diffusion of heat and metals among SPH particles \citep{wadsley08}
have been shown to reproduce quantitatively the properties of the circum-galactic medium of galaxies  at $z\sim3$ \citep{shen11,shen12}, which reflect the complex exchange 
of matter, energy, and metals between galaxies and their surroundings.

%%%%%%%%%%%%%%%%%%%%%%%%%%%%%%%%%%
% PSEUDOBULGE DEFINITION
%%%%%%%%%%%%%%%%%%%%%%%%%%%%%%%%%%
\section{Defining the Pseudobulge} \label{definition}

%The definition of ``pseudobulge" must be taken with caution here, since in Eris such component is not easily separable from the disk in terms of formation path. 
Throughout this paper, the use of the term ``pseudobulge" is dictated by the low S\'ersic index of the inner regions of the galaxy \citep{kormendy04}. 

Before formally decomposing the galaxy in its main components, it is instructive to take a qualitative look at the Eris galaxy and its inner regions across cosmic time. Figure \ref{evolution} shows the photometric evolution of Eris (left panel) and its inner regions (right panel) in face-on and edge-on portraits. It is these inner regions that we 
identify as the ``pseudobulge", without trying to separate out the bar-like component of the potential.  On the left panels we show composite rest-frame $i$,V, and far UV 
images of Eris from $z=4$ to $z=0$ generated by  {\tt SUNRISE} \citep{jonsson06}, and on the right we show iso-density contours of the stellar distribution within 5 kpc of the center. The disk-like 
structure that characterizes the pseudobulge is present already at $z=4$, and the stellar disk grows around this initial structure as time goes on. A stellar bar is present at this redshift, and is destroyed at $z=3$, an epoch characterized by a series of minor mergers (see left panel). Between redshifts 2 and 1 the bar reforms, but by $z=0$ the bar is reduced to a very small nuclear bar only a few
softening lengths in size, and has left behind a flattened spheroidal pseudobulge (see Figure \ref{evolution}).

In order to identify a subset of particles that are representative of the $z=0$ pseudobulge, we decompose the galaxy based on its photometric and kinematic properties by selecting particles that (i) lie in the region where the light profile is dominated by the S\'ersic profile, and (ii) are not kinematically cold. 

\begin{figure}[t!]
\begin{center}
\includegraphics[scale=0.7]{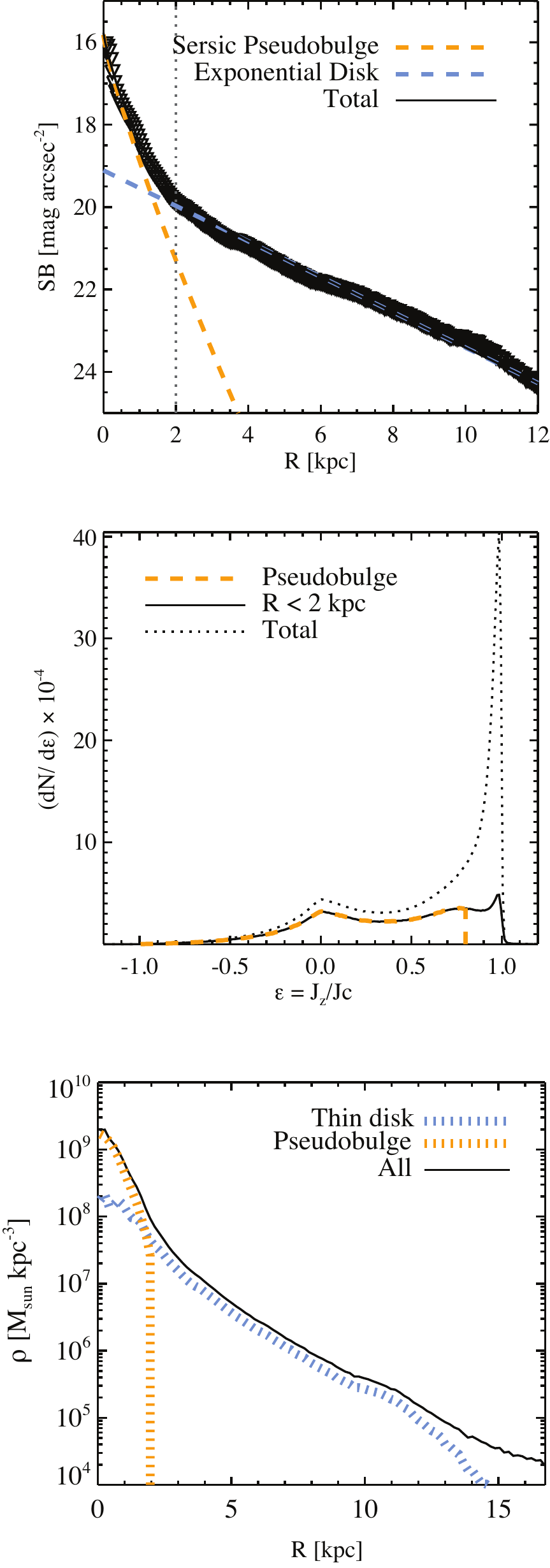}
\caption{Photometric and kinematic definition of Eris' pseudobulge at $z=0$. {\it Top panel}: Eris' surface brightness profile in the $i$-band ({\it black points}), and 
its best-fit decomposition into S\'ersic bulge and exponential disk given by {\tt GALFIT}. The dotted line shows the radius below which the pseudobulge component 
dominates the light profile. {\it Middle panel}: Angular momentum distribution of all stars ({\it dotted line}), of stars located within $3 R_e = 2$ kpc ({\it solid 
black line}), and of the pseudobulge stars with $R<3 R_e$ and $\epsilon\equiv J_z/J_c<0.8$. {\it Bottom panel}: Surface density profile of the 
kinematically decomposed pseudobulge ({\it orange dotted line}) and thin stellar disk ({\it blue dotted line}), as well as of all star particles within 20 kpc from the center.}
\label{pseudobulge}
\vspace{+0.1cm}
\end{center}
\end{figure}

Figure \ref{pseudobulge} (top) shows the $i$-band surface brightness profile of the galaxy at $z=0$ and the photometric decomposition obtained with  the 
{\tt GALFIT} software \citep{peng02}.The resulting parameters are: an integrated bulge magnitude $M_i=-19.3$, a $B/D= 0.35$, a S\'ersic index $n=1.4$ projected face-on, an effective 
radius $R_e = 0.68$ kpc, and an exponential disk scale length $R_s = 2.5$ kpc.  The pseudobulge dominates the light profile within a radius of $\sim 2$ kpc, or three 
effective radii. To isolate the warmer pseudobulge component and remove stars that are kinematically cold (and most likely belong to the thin disk), we decompose the star 
particles kinematically according to their angular momentum \citep{abadi03,scannapieco09} as follows.  We define the quantity $\epsilon=J_z/J_c$, where $J_z=J_z(E)$ is the 
angular momentum of a star of total energy E along the $z$ direction; $\epsilon$ is calculated on a centered and aligned frame of reference, where the $xy$ plane coincides with the plane 
of the galactic disk. The maximum angular momentum of a particle of total energy $E$ is given by $J_c = V_c*r$, where $V_c = \sqrt{GM(<r)/r}$ is the circular velocity. 
In this decomposition scheme, stars with low angular momentum occupy the region around $\epsilon=0$ and belong to the pseudobulge and stellar halo, while stars with
$\epsilon\sim1$ have nearly circular orbits and belong to the thin, centrifugally-supported disk. 

\begin{figure*}[ht!]
\begin{center}
\includegraphics[scale=0.6]{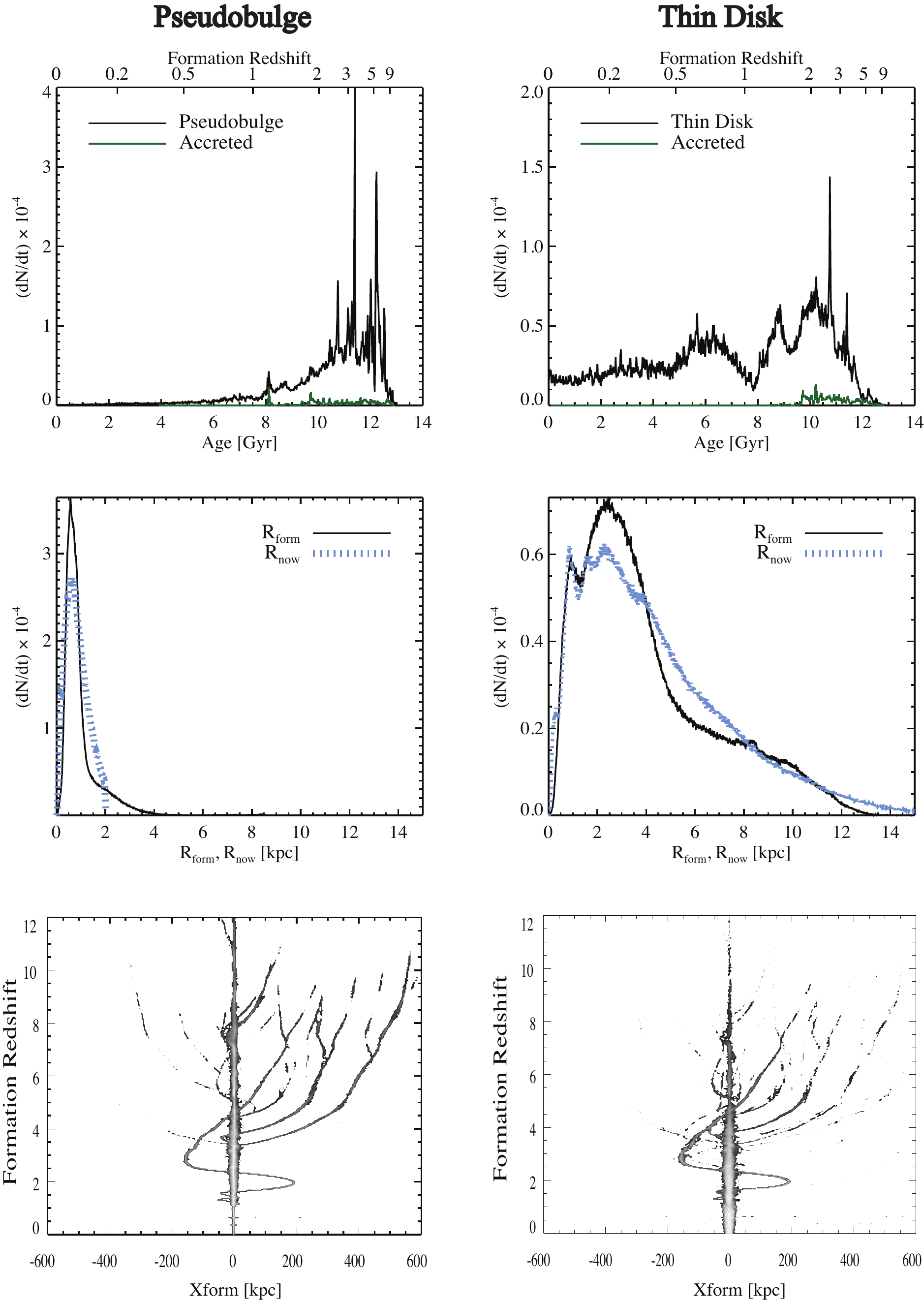}
\caption{Origin of pseudobulge stars. {\it Top panels}: Stellar age distribution of the present-day pseudobulge ({\it left}) and thin disk ({\it right}).
Star particles that have been accreted (``non in-situ star formation")  are shown with the green curve. {\it Middle panels}: Formation radius ({\it black solid line}) and current radius 
({dotted blue line}) of 
pseudobulge ({\it left}) and thin disk ({\it right}) stars. {\it Bottom panels}: Merger history of the pseudobulge ({\it left}) and thin disk ({\it right}). The $y$-axis shows the formation
redshift, the $x$-axis the formation coordinate in the frame of the host galaxy.}
\label{origin}
\vspace{+0.1cm}
\end{center}
\end{figure*}

The resulting stellar angular momentum distribution is shown in the middle panel of Figure \ref{pseudobulge}. The dotted line shows the distribution of all particles within 16 
kpc from the center of the galaxy, showing the prominence of the thin stellar disk at $\epsilon \sim 1$ and warmer components at lower values of $\epsilon$.  It is clear that there is 
a continuum in the angular momentum distribution going from the disk to the pseudobulge and stellar halo (the peak at $\epsilon=0$). This is consistent with the fact that the central 
pseudobulge is effectively the low angular momentum tail of the disk rather than a truly distinct component \citep[see also][]{Dubois12}. The solid line shows the angular momentum 
distribution of all particles within $r<2$ kpc, and the dashed yellow line shows the same distribution after applying an angular momentum cut at $\epsilon<0.8$, which excludes thin 
disk particles. Thus, in our definition, the pseudobulge spans a wide range of angular momenta, having a hot component at $\epsilon \sim 0$ but also a tepid, disk-like component at  intermediate angular momenta. The surface density profile of the kinematically derived pseudobulge and thin disk are shown on the bottom panel of figure \ref{pseudobulge}. 

In summary, the pseudobulge is defined as consisting of particles star particles that lie within a radius $r<2$ kpc and are kinematically warm with $\epsilon<0.8$. 
Particles with  $\epsilon \geq 0.8$ are assigned to the thin disk. 
%Particles kinematically decomposed in this manner are considered tracers of the pseudobulge and thin disk, and shall not be taken as unique representations of these components. 

%%%%%%%%%%%%%%%%%%%%%%%%%%%%%%%%%%
% FORMATION AND EVOLUTION OF THE PSEUDOBULGE
%%%%%%%%%%%%%%%%%%%%%%%%%%%%%%%%%%
\section{FORMATION AND EVOLUTION OF THE PSEUDOBULGE}

Having defined a subset of particles that belong to the present-day pseudobulge, we proceed to study its formation and evolution. 
%%%%%%%%%%%%%%%%%%%%%%%%%%%%%%%%%%
% ORIGIN
%%%%%%%%%%%%%%%%%%%%%%%%%%%%%%%%%%
\subsection{Origin}

The main mechanism for the assembly of Eris' pseudobulge is by no means secular and it is not distinct from the formation path of the inner disk. Figure \ref{origin} shows the distributions of age and formation radius, and the merger history of the present-day pseudobulge and thin disk.  Pseudobulge stars form in an early burst at $z\sim3-5$,  while thin disk stars have a spread of ages. Accreted stars (green histogram) are identified as particles that were born in a subhalo other than the main progenitor.  In order to track the origin of these particles, we use the {\tt AMIGA Halo Finder} \citep[AHF,][]{gill04,kollmann09} to identify the host galaxy and subhalos from $z=10$ onwards. We find that only 4\% the pseudobulge mass was accreted, and that the bulk of this component formed in situ. Similarly, 98\% of the thin disk's mass formed in situ. 

The middle panel shows the distribution of formation radius $R_{\rm form}$ and present-day radii $R_{\rm now}$ of the pseudobulge (left panel) and the thin disk (right panel) that formed in-situ. Most of the pseudobulge stars formed within the inner 2 kpc of the center of the galaxy, and inward migration of stars in the region between 2 and 4 kpc contributes only 13\% of its final mass. The formation of the thin disk occurs also mostly in-situ, although 11\% of all stars that formed within $r=4$ kpc migrate outwards to 15 kpc, possibly via the resonant scattering of stars by transient spiral arms \citep{roskar08}. 

\begin{figure}
\centering
\includegraphics[width=0.45\textwidth]{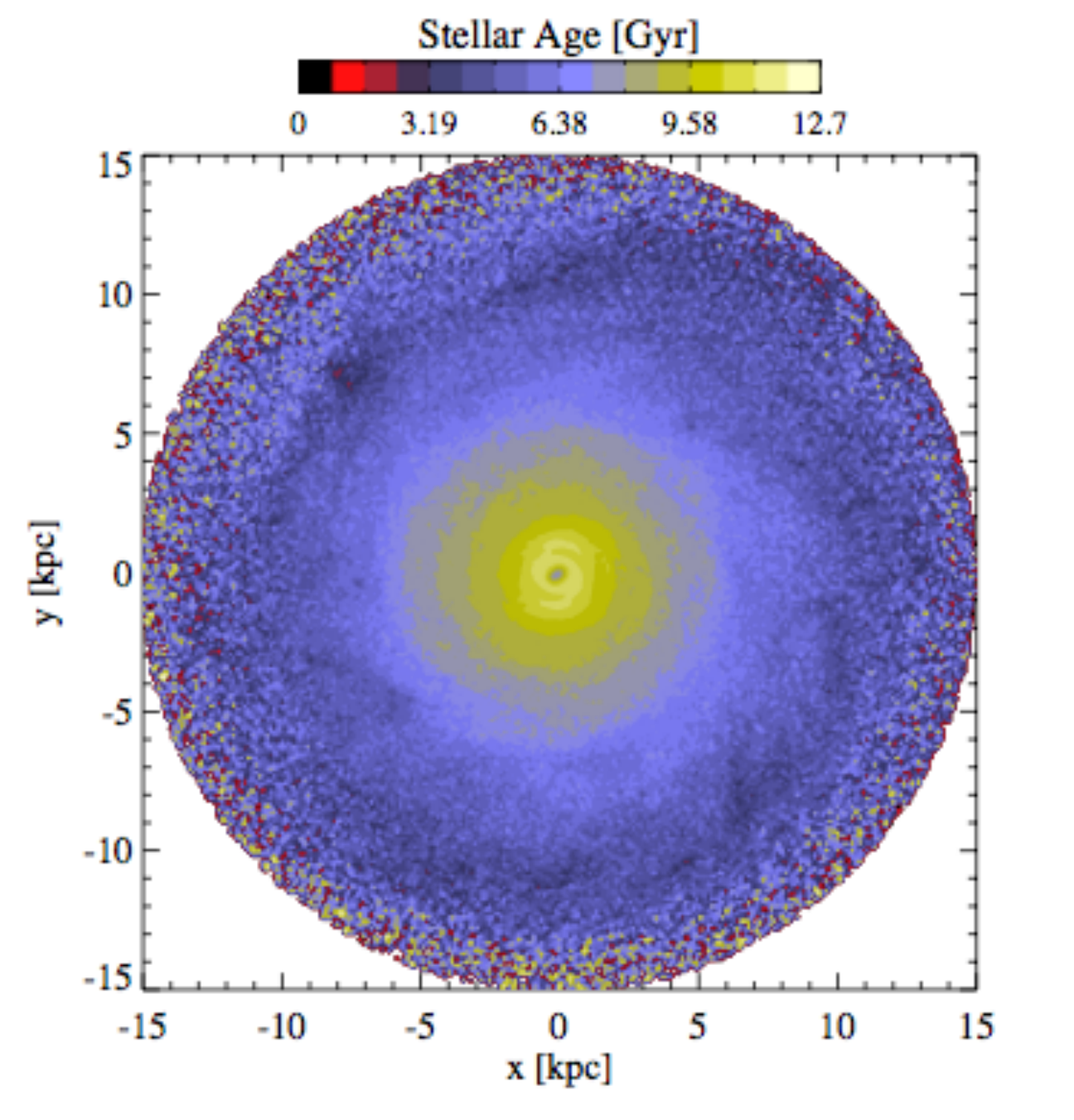}
\caption{Map of the mass-weighted age distribution of the galaxy projected face-on through a 2 kpc slice. In terms of stellar age, the transition from the pseudobulge to the inner disk is smooth. Some key features are the spiral spiral structure contained within the pseudobulge and a weak, younger nuclear bar within it.}
\label{age_map}
\vspace{+0.5cm}
\end{figure}

The formation histories of the pseudobulge and the thin disk (bottom panels) are remarkably similar, in the sense that mergers do not contribute to the formation of the pseudobulge more than they contribute to the formation of the thin disk. In our simulations then, the pseudobulge is the building block of the rest of the galactic disk, and does not form \textit{a posteriori} from instabilities in it. As a result, the age distribution in the plane of the disk is a continuum from the inner to the outer regions, as shown in Figure \ref{age_map}. 
 
\begin{figure}[t!]
\begin{center}
\includegraphics[width=0.45\textwidth]{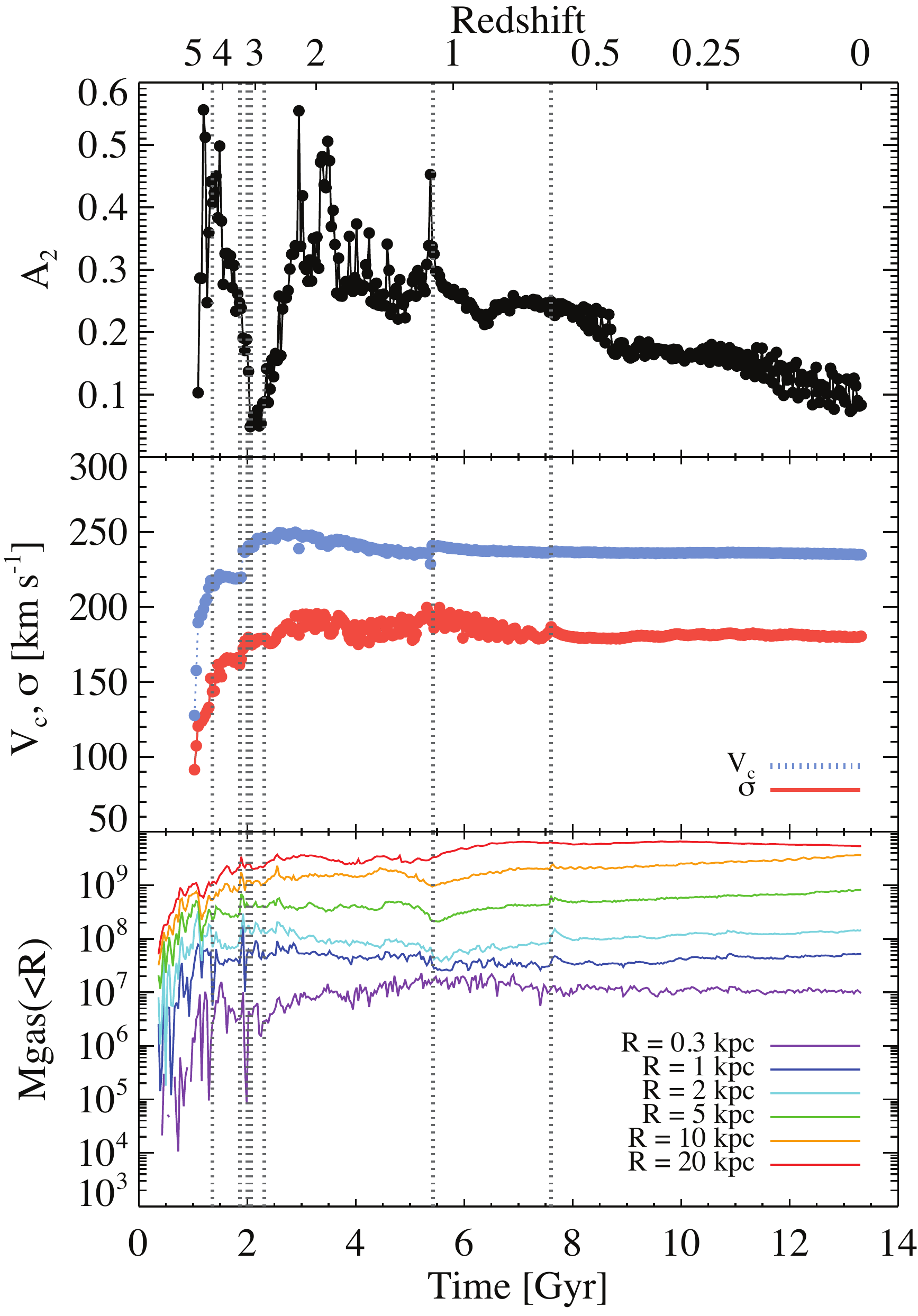}
\caption{Top panel: evolution of the amplitude of the m=2 Fourier mode of the stellar density distribution. Middle panel: Evolution of the circular velocity and the velocity dispersion within 2 comoving kpc from the center. Bottom panel: Evolution of the gas mass enclosed within different radii as a function of redshift. Large inflows followed by outflows occur during minor mergers at $z>4$, at $z=3$ and $z=1$. Dotted gray lines represent minor merger events at $z<5$.}
\label{mgas_m2_evol}
\end{center}
\end{figure}

%%%%%%%%%%%%%%%%%%%%%%%%%%%%%%%%%%
% EVOLUTION
%%%%%%%%%%%%%%%%%%%%%%%%%%%%%%%%%%
\subsection{Evolution}

While in isolated galaxy models a massive stellar disk is in place in the initial conditions, in hierarchical galaxy formation the disk forms and grows over time, is at least partially destroyed during mergers, re-grows after them, and suffers continuing external perturbations by infalling matter, in particular satellites. All such phenomena  play an important role in redistributing mass and angular momentum in the disk, contributing to the formation of a central, more concentrated component that we can recognize as the ``bulge". At late times, when the galaxy is no longer harassed by mergers, less violent (although not strictly secular) processes begin to play an important role in the evolution of the galaxy. 

Secular evolution is not a single mechanism, but rather a collection of different phenomena that can in principle take place even on 
dynamical timescales as a result of rapidly changing conditions both internal and external to the disk, as expected in the
hierarchical assembly of galaxies. These processes
have been studied predominantly with isolated galaxy simulations \citep{debattista06}. The formation of a stellar bar is the first step in the process. The bar can then buckle into a kinematically hot spheroid as a result of bending modes, and drive an inflow of gas, resulting in new star formation taking place place in the center. The strength of the bar and therefore the effectiveness of buckling and inflows depend on the relative proportions of gas and stars, with more gas tending to weaken the bar \citep{hasan93} and preventing buckling to first order \citep{debattista06}. Both the reshaping of pre-existing stars by internal instabilities and the new central star formation can contribute to the build-up of the bulge from disk material.  The Eris simulation has a mass and force resolution comparable to some of the reference studies of bar instabilities and secular evolution published in the literature \citep{debattista06, athanassoula08} despite the added complexity of the cosmological context. It thus allows to address in detail the role of these different processes traditionally
considered part of secular evolution.

%%%%%%%%%%%%%%%%%%%%%%%%%%%%%%%%%%
% THE ROLE OF MERGERS
%%%%%%%%%%%%%%%%%%%%%%%%%%%%%%%%%%
\subsubsection{The Role of Minor Mergers}

Although only a small fraction of the mass of the pseudobulge is contributed by accreted satellites, minor mergers aid to its evolution by destroying the bar and redistributing angular momentum in the disk. 

Figure \ref{evolution} shows that a star forming disk-like structure with a strong bar mode is already in place at $z=4$ and evolves significantly afterwards. 
 In order to quantify the strength of the bar as a function of redshift, we calculate the amplitude of the $m$ Fourier mode of the stellar density distribution in polar coordinates $ \Sigma(R,\theta,t)$ at each simulation output as follows:
 \begin{equation}
 A_m (R,t)= \frac{1}{\Sigma(R,t)}  \left |  \int_{0}^{2\pi} \Sigma(R,\theta',t)\, e^{-i m \theta'} d\theta'\, \right |.
 \end{equation}
The phase is given by
\begin{equation}
\phi(R,t) = \tan^{-1}\left [ \frac{\Re(A_m)}{\Im(A_m)}\right ]
\end{equation}
 where $\Re(A_m)$ and $\Im(A_m)$ are the real and imaginary components of $A_m$, respectively. When a strong bar is present, the phase of the $m=2$ remains constant over the radial extent of the bar  $R_{\rm bar}$, and changes drastically beyond it. Therefore the strength of the bar $A_2$ can be measured as the maximum amplitude within a radius $R_{\rm bar}$. In cases where the bar is weak, we measure $A_2$ and the maximum amplitude within a fixed radius $R=2$ kpc. The top panel of Figure \ref{mgas_m2_evol} shows the evolution of $A_2$ as a function of redshift. Several minor mergers (shown as dotted vertical lines) at $z\sim3$ destroy the bar, which reforms quickly by $z=2$ and survives for 1 Gyr. After the last minor merger at $z=1$ the bar steadily weakens, becoming negligible at $z=0$. 
      
It is important to stress that a bar forms as soon as the inner disk forms at high redshift ($z > $4), as demonstrated by the amplitude 
of the $m=2$ mode at those early epochs. 
The disk therefore develops a concentrated inner disk (a ``bulge") with a low S\'ersic index as typical in present-day pseudobulges already from the very beginning and on fast, dynamical timescales. Such concentrated inner disk, which coincides with the bar, undergoes continued evolution (Figures \ref{evolution} and \ref{mgas_m2_evol}).  What we witness is the action of the processes normally included
in secular evolution but happening in a dynamical fashion; indeed there are continued external tidal triggers, 
resulting in a sequence of repeated, fast developing non-axisymmetric
instabilities interrupted by minor mergers rather than a slow growth of non-axisymmetric modes giving rise to a 
bulge-like component only after several billions of years.

  \begin{figure}[t!]
\includegraphics[width=0.45\textwidth]{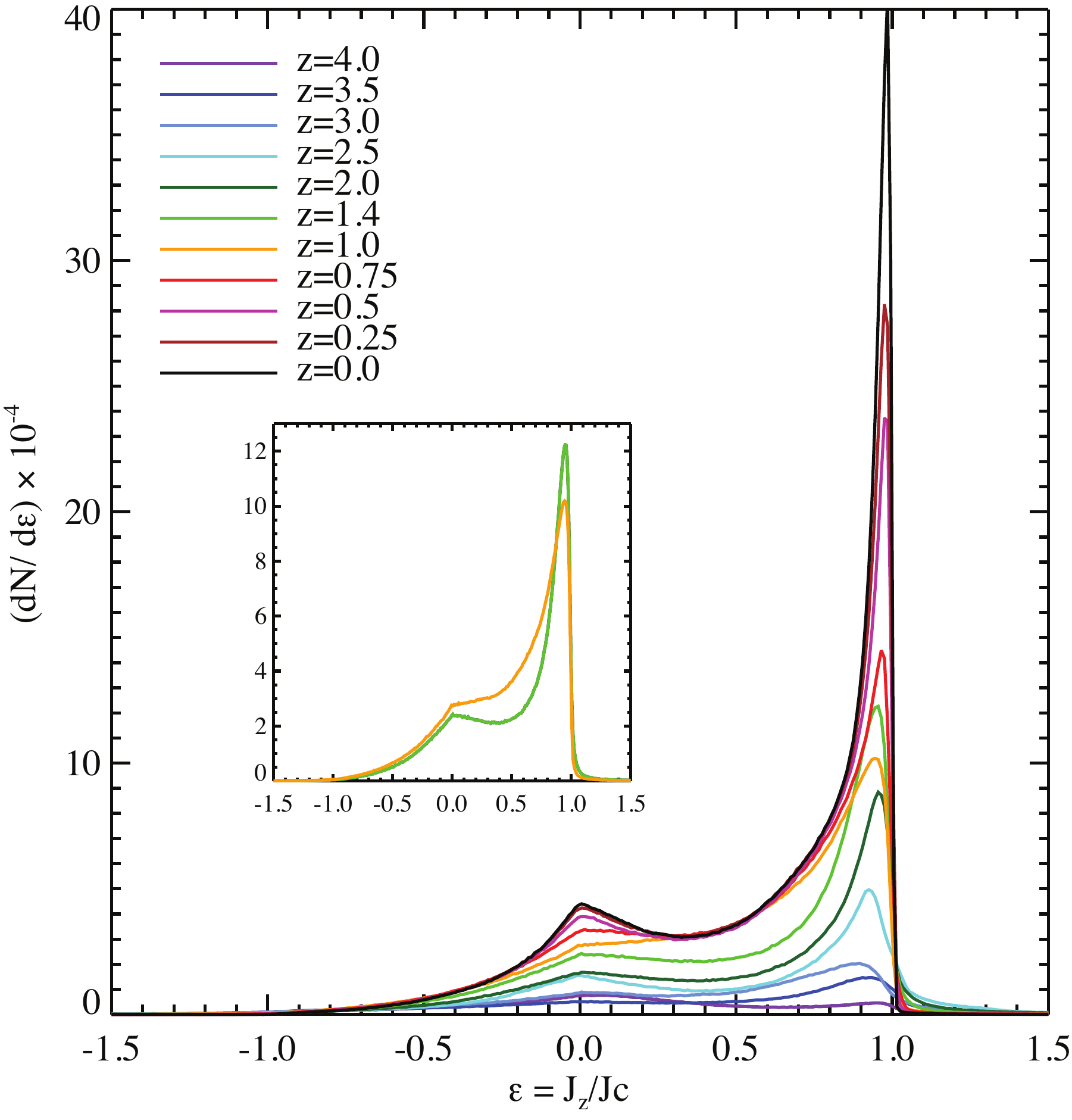}
\caption{Evolution of the $\epsilon=J_z/J_c$ distribution as a function of redshift for all stars within the optical radius. Disk stars populate the region around $\epsilon=1$ (co-rotation) while stars that occupy the regions around $\epsilon=0$ form part of the spheroid, which includes pseudobulge and halo stars. Particles with intermediate angular momentum can be associated with either the pseudobulge or  the thick disk according to their binding energies. The first structure to form has low angular momentum and dominates the distribution down to $z\sim3$, at which point the disk component has grown massive enough to become dominant.}
\label{jzjc_evol}
\vspace{+0.1cm}
\end{figure}

\begin{figure}
\begin{center}
\vspace{+0.3cm}
\includegraphics[width=.35\textwidth]{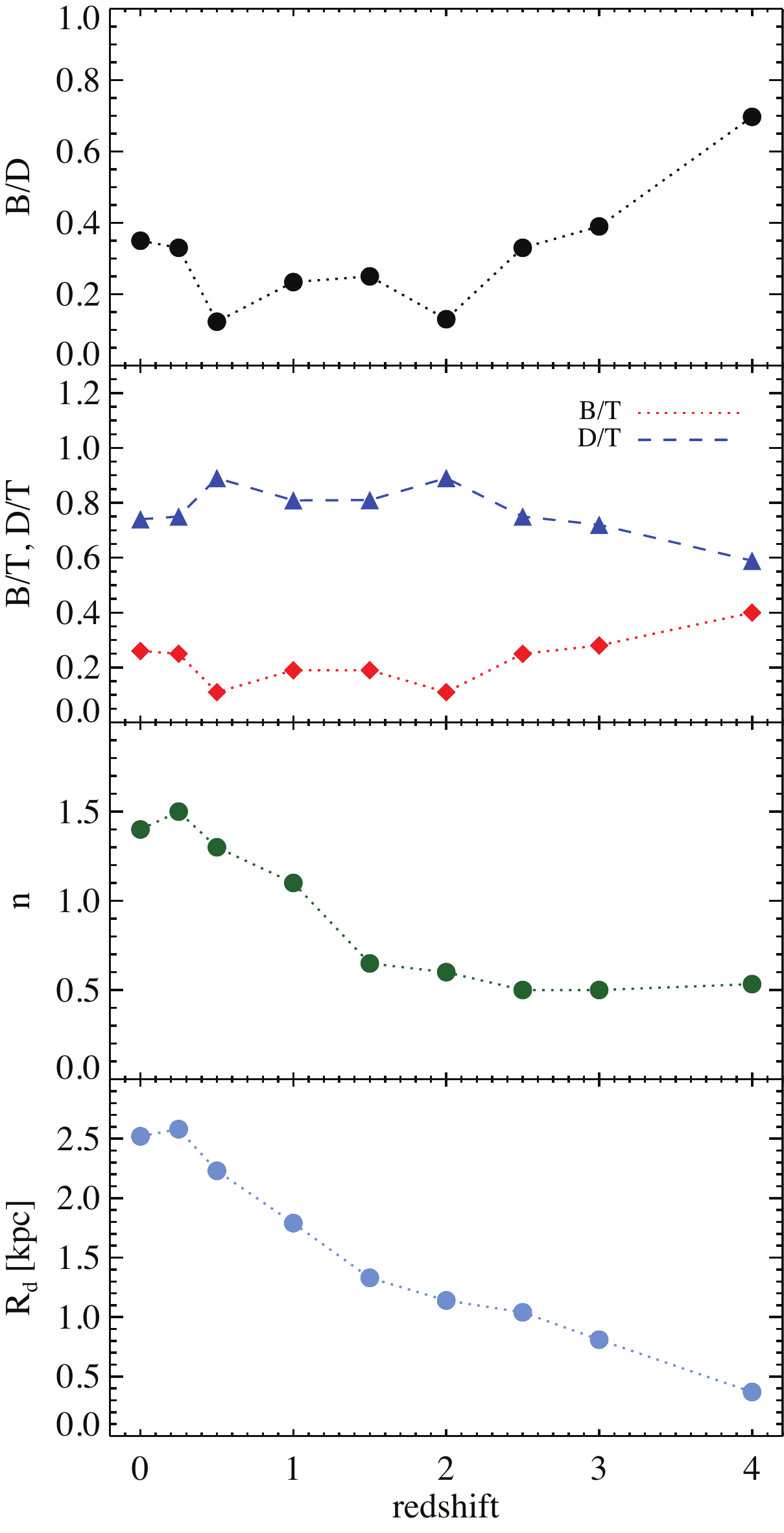}
\caption{Evolution of the photometric properties of the galaxy as function of redshift. The top panels show the evolution of the photometric B/D, B/T and D/T ratios obtained by fitting a S\'ersic profile to the bulge and an exponential profile to the disk with {\tt Galfit} \citep{peng02}. The bulge component includes the contribution of the  bar).  The evolution is non-monotonic since $z=4$. The second panel shows the evolution of the S\'ersic index $n$. At early times the bulge has a Gaussian-like profile ($n\sim0.5$) which steepens at late times to reach a maximum value of only $n\sim1.5$, suggesting that there is never a classical bulge in Eris. The bottom panel shows the growth of the scale length of Eris' disk. There is a remarkable similitude in the growth rate of the S\'ersic index and the scale length, suggesting that the formation of the disk and the pseudobulge are tightly interconnected.}
\label{photometry_evol}
\end{center}
\vspace{+0.1cm}
\end{figure}

The evolution of the average velocity dispersion, $\sigma$, of the central regions (Figure \ref{mgas_m2_evol}) is tightly interconnected to the merging history of the galaxy and the evolution of the bar. As the protogalaxy begins to form and the bar is violently destroyed by minor mergers, the velocity dispersion increases, reaching a peak of $200\kms$. After $z=3$ the central velocity dispersion is only marginally affected by mergers and remains nearly constant from $z=0.6$, to the present time. This is because late mergers are more strongly tidally disrupted by the disk and most of the remnant material does not become part of the pseudobulge, whereas at early times the bulge dominates the protogalaxy are is more vulnerable to mergers. 

Thus, the weakening of the bar at $z<1$ is not accompanied by an increase in velocity dispersion as would be expected if the buckling instability was the major driver of the formation of the pseudobulge. This is consistent with studies that show that the buckling instability requires long-lived strong bars coupled with high central stellar densities to ensue, whereas in Eris the bar weakens and becomes more compact at low redshift due to angular momentum losses via dynamical friction and bar-dark matter halo interactions. 

In addition to increasing the velocity dispersion, minor mergers also redistribute angular momentum within the disk. Figure \ref{jzjc_evol} shows the distribution of angular momentum of all particles within the optical as a function of redshift. The first structure to form is composed of mostly low angular momentum material. By $z\sim3.5$ the disk begins to dominate the distribution. The disk and spheroid grow together, except at $z=1$ when a minor merger heats up the inner regions, destroys the bar, and redistributes angular momentum from high to intermediate values. The inset shows a detail of this angular momentum redistribution before (green) and after (orange) the minor merger, showing that there is a transfer of angular momentum from the disk ($\epsilon\sim1$) to the pseudobulge and thick disk  ($\epsilon<0.8$).

%%%%%%%%%%%%%%%%%%%%%%%%%%%%%%%%%%
% GAS INFLOW
%%%%%%%%%%%%%%%%%%%%%%%%%%%%%%%%%%
  \subsubsection{The Role of Gas Inflow}
 
\begin{figure}[t!]
\vspace{+0.3cm}
\includegraphics[width=0.5\textwidth]{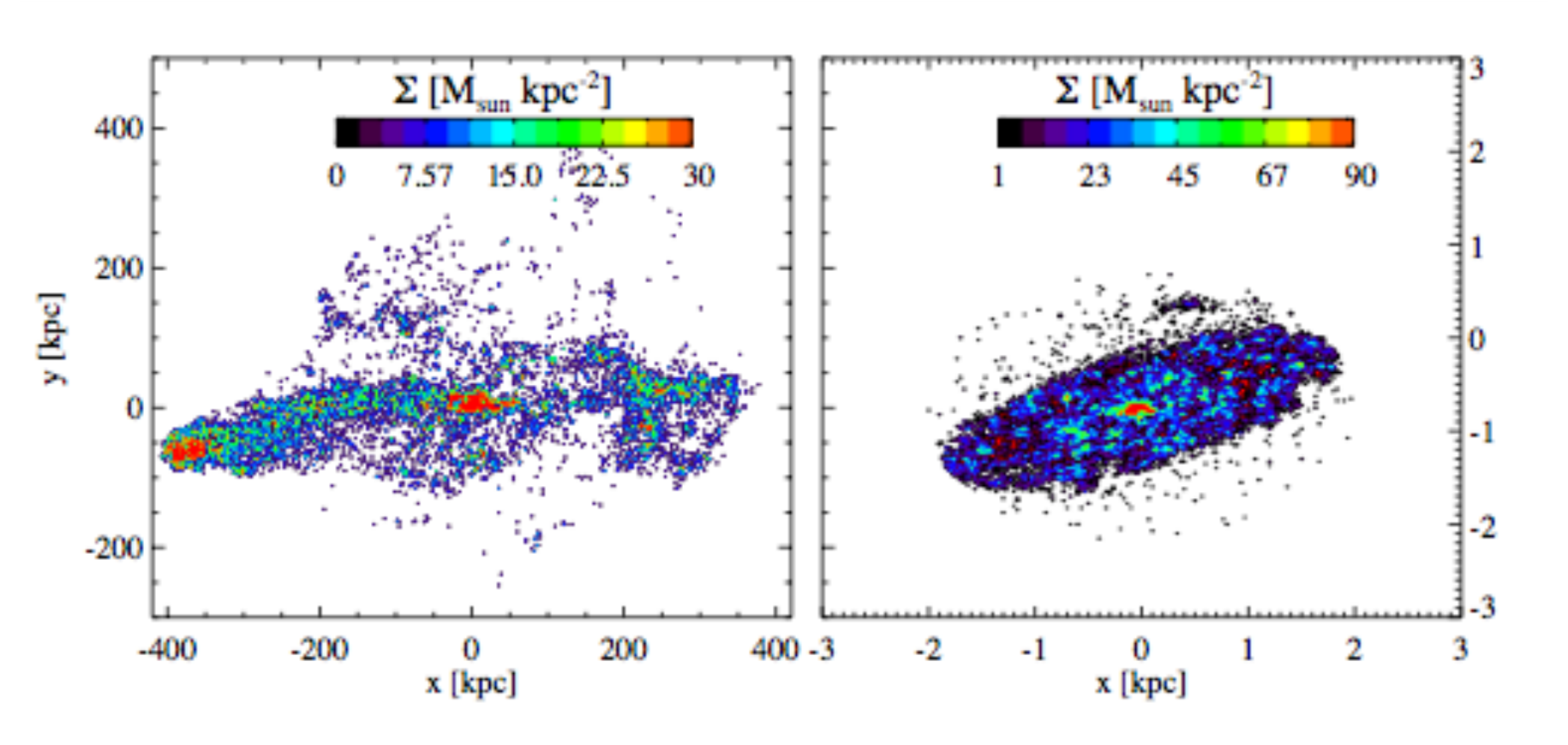}
\caption{Left: Surface mass density map of the gas within 5 kpc of the center of Eris at $z=0$.  Right: Origin of the inner gas disk; surface mass density distribution of the gas particles identified to be in the region $r<2$ kpc at $z=0$ traced at $z=1$. 95\% of the gas that today is in the inner 5 kpc was at $r>50$ kpc at $z=1$.}
\label{inflows}
\vspace{+0.3cm}
\end{figure}

 In Eris, there is steady inflow of  cold gas at $z \leq1$, that originates from accreting cold flows entering the halo and feeding the galaxy, and continues 
within 
the galaxy itself down to its nucleus.
It contributes to the assembly of the outer disk, to replenishing the gas reservoir at smaller radii, and to sustaining a rate of star formation of 1-2 $\sfr$ during 
this time. Figure \ref{mgas_m2_evol}  (bottom panel) shows  the total gas mass enclosed within 0.3 to 20 kpc. While the gas mass within the galactic disk only 
increases by a factor of $\sim2$ from redshift one to zero, the steady inflow of gas contributes to the weakening of the stellar bar suggested by the 
decreasing power of the $m=2$ mode (top panel) by increasing the central baryonic concentration thus destabilizing
box orbits, and contributes to the build up 
of the pseudobulge with new star formation but without increasing its stellar velocity 
dispersion. The fact that the  inflow of gas has a cosmological origin is shown in Figure \ref{inflows}, where we have traced back the gas that is at $r<2$ kpc today 
(right panel) 
to its location at $z=1$ (left panel). Gas that is located today within 2 kpc from the center, was funneled to the center via and dense cold gaseous filaments at high redshift. Minor merger events, shown as dashed lines in Figure \ref{mgas_m2_evol}, with peaks in the gas mass distribution and followed by outflows, do not contribute significantly to the gas mass accretion. 

%%%%%%%%%%%%%%%%%%%%%%%%%%%%%%%%%%
% PHOTOMETRIC EVOLUTION
%%%%%%%%%%%%%%%%%%%%%%%%%%%%%%%%%%
\subsubsection{Photometric Evolution of the Bulge and the Disk}

Figure \ref{photometry_evol} shows the evolution of the photometric properties of the galaxy, obtained by running {\tt GALFIT} on rest-frame $i$-band images produced 
by {\tt SUNRISE} in a range of redshifts. Each image is fit with S\'ersic index and an exponential disk profile, and we do not attempt to decompose the bar 
separately. The top panels show the evolution of the B/D, B/T, and D/T ratios. These ratios do not evolve monotonically; their value is dictated by the evolution of 
the disk, the bulge, and the bar. As shown in the kinematic decomposition (Figure \ref{jzjc_evol}), the bulge component forms first and dominates the light 
distribution until $z\sim3$. The B/D ratio fluctuates in response to the strength of the bar, and at $z<1$, the B/D ratio and the S\'ersic index $n$ increase  as 
the large-scale bar gradually dissolves leaving behind a fairly compact 
pseudobulge with a nuclear bar. Since the bar forms dynamically immediately after the disk arises, the $B/D$ of the galaxy is within a factor of 2-3 from the final 
value since the  beginning with only rare instances, such as after the last major mergers, in which the bar/pseudobulge are temporarily destroyed. Instead of 
increasing the central cusp, mergers have the opposite effect here, at least at high redshift, where they destroy the stellar bar effectively lowering the photometric bulge-to-disk ratio.  
The scale length of the disk, $R_d$ doubles since $z=2$ due to a steady gas inflow that builds the disk inside-out. Although the pseudobulge forms during an early 
burst of star formation, the S\'ersic index of the inner region of the galaxy is never high, implying that at birth the pseudobulge stars are kinematically tepid, not hot.

%%%%%%%%%%%%%%%%%%%%%%%%%%%%%%%%%%
% SUMMARY AND DISCUSSION
%%%%%%%%%%%%%%%%%%%%%%%%%%%%%%%%%%
\section{Summary and Discussion}

We have studied bulge formation in a late-type disk galaxy. The bulge that forms has all the characteristics of a pseudobulge, consistent with the fact that late-type galaxies, including the Milky Way, have pseudobulges. However, while it is normally believed that such bulges are the result of a slow growth from internal dynamical instabilities, termed ``secular evolution", we have shown that in our simulations the process is fast and dynamical. This is in very good agreement with observations of pseudobulges in Milky Way-like galaxies in the local universe \citep{carollo2007}. Bars develop on a dynamical timescale as soon as the disk begins to assemble. This, and not fragmentation into clumps, seems to be the main manifestation of disk self-gravity and internal dynamical evolution at the galaxy masses considered here. The recent discoveries of clumpy galaxies at high z, and the simulations developed to explain them, are compatible with this picture as long as one accepts that disk fragmentation becomes a mode of internal evolution only at much larger disk mass, presumably corresponding to a halo mass in excess of $10^{12} M_{\odot}$, as the Toomre parameter is easily brought below unity by accretion of gas and disturbances \citep[e.g.][]{agertz09, genel10}.

The pseudobulge is part of the low angular momentum tail of the disk material rather than being a separate component, not surprisingly since it is disk material that was assembled at early times in the initial disk assembly phase (due to early instabilities) or that was brought to the center by gas inflows at later times. Therefore there is a age distribution of stars in the pseudobulge, which is not young but rather contains a substantial old component, contrary to common objections against the internal instability/secular evolution picture of bulge formation.  If this scenario is correct, progenitors of late-type disks, which are the majority of present-day galaxies, are thus expected to have strong bars and small ``photometric pseudobulges"  (i.e. indicated by low S\'ersic index inner stellar profiles) already at $z > 3$. Such bars are a significant fraction of the galaxy size in length, more than in the present-day galaxies. At masses significantly lower than those considered here, supernovae feedback might maintain a larger gas fraction and prevent the formation of a massive, bar unstable stellar disk until much later times, producing younger pseudobulges \citep{brook10}, or no bulge at all in dwarf galaxies \citep{governato10}.\\ 

Our results can be summarized as follows:
\begin{itemize}

\item In our simulations, the bulk of the pseudobulge's mass forms early ($z\sim4$), in situ, and rapidly (2 Gyr) from non-axisymmetric instabilities such
as bars, not secularly. While the disk undergoes several episodes of such instabilities it never fragments into massive clumps as proposed in other scenarios
of early bulge formation.
Minor mergers contribute only 4\% of its final mass. 

\item The formation path of the pseudobulge and of the thin disk are indistinguishable in terms of merger history, suggesting that the inner disk and the pseudobulge are not distinct components. And since in our simulations the age profile distribution is continuous from the inner to the outer regions, the pseudobulge is simply the building block of the galaxy and of the inside-out disk formation scheme. 

\item We find that the evolution of the pseudobulge is tightly intertwined with the evolution of the bar. Present-day pseudobulge particles formed early in a bar-like 
configuration. The bar mode was destroyed at $z\sim3$ by minor mergers, but reforms quickly by $z\sim2$. The steady increase in gas mass due to the presence of the 
bar and the redistribution of angular momentum within the stellar distribution at $z\sim1$ lead to the gradual dissolution of the bar, leaving behind a compact
flattened bulge and nuclear bar, and a 
steady increase of the  S\'ersic index from $n=0.5$ at $z=4$ to $n=1.4$ at the present epoch. 

\end{itemize}

Finally, we note that the pseudobulge formation process shown here is not peculiar to our simulation code and star formation prescription and parameters. Recent work 
by \cite{Okamoto12} shows that pseudobulges that form in galaxies of the mass of Eris and higher form via similar mechanisms as shown here in simulations of only a 
factor of $\sim2$ lower resolution, lower star formation density threshold, and strong kinetic feedback giving rise to hydrodynamically
decoupled winds. The early and rapid formation  of galactic bulges in 
cosmological simulations suggests that attempts to produce massive bulgeless galaxies, such as those that seem to dominate in the Local Volume \citep{kormendy10},
should focus on suppressing this early burst of star formation by introducing, for instance, an additional mode of early stellar feedback \cite{brook12}.

\acknowledgements
We acknowledge useful discussions with Anna Cibinel, Sandra Faber, and Alvio Renzini. Support for this work was provided by the NSF through grant OIA-1124453 and NASA through grant NNX12AF87G (P.M.), and by the ETH Zurich Postdoctoral Fellowship and the Marie Curie Actions for People COFUND Program (J.G.). The Eris Simulation was performed at NASA's Pleiades supercomputer, the UCSC Pleiades cluster, and the Swiss CSCS's ROSA Cray XT-5.  

\bibliographystyle{apj}

\end{document}